# FINEMBEDDIFF: A COST-EFFECTIVE APPROACH OF CLASSIFYING FINANCIAL DOCUMENTS WITH VECTOR SAMPLING USING MULTI-MODAL EMBEDDING MODELS

**Anjanava Biswas[*1], Wrick Talukdar[*2]**

[*1,2]AWS AI & ML, IEEE CIS, India.

DOI : https://www.doi.org/10.56726/IRJMETS57269

## ABSTRACT

Accurate classification of multi-modal financial documents, containing text, tables, charts, and images, is crucial but challenging. Traditional text-based approaches often fail to capture the complex multi-modal nature of these documents. We propose FinEmbedDiff, a cost-effective vector sampling method that leverages pre-trained multi-modal embedding models to classify financial documents. Our approach generates multi-modal embedding vectors for documents, and compares new documents with pre-computed class embeddings using vector similarity measures. Evaluated on a large dataset, FinEmbedDiff achieves competitive classification accuracy compared to state-of-the-art baselines while significantly reducing computational costs. The method exhibits strong generalization capabilities, making it a practical and scalable solution for real-world financial applications.

## I. INTRODUCTION

The financial industry generates and processes vast amounts of multi-modal data, encompassing reports, prospectuses, regulatory filings, and various other document types that combine textual content with tables, charts, and images. Accurate and efficient classification of these multi-modal financial documents is crucial for numerous applications, including risk assessment, compliance monitoring, investment analysis, and decision-making processes. However, the multi-modal nature of these documents poses significant challenges for traditional text-based classification methods, which often fail to capture the rich semantic information present in both textual and visual components.

While recent advances in deep learning have led to the development of powerful multi-modal models, their practical application in financial document classification remains limited. Existing multi-modal approaches often require extensive computational resources for end-to-end training, rendering them impractical for large-scale deployments or resource-constrained environments. Furthermore, these models may struggle to effectively capture the intricate relationships between textual and visual components, which can vary substantially across different financial document types and domains.

To address these challenges, we propose FinEmbedDiff, a cost-effective vector sampling method for classifying multi-modal financial documents using pre-trained multi-modal embedding models. Our approach leverages these pre-trained models to generate rich multi-modal embedding vectors that capture the semantic information from both textual and visual components of financial documents. To classify new documents, we create their multi-modal embedding vectors using the same pre-trained models, and compare them with a set of pre-computed class embeddings using vector similarity measures, such as cosine similarity or L2 distance.

The key advantages of our FinEmbedDiff method are:

1. **Cost-effectiveness**: By leveraging pre-trained multi-modal embedding models and avoiding computationally expensive end-to-end training, our approach significantly reduces computational costs and resource requirements, making it practical for large-scale deployments.
2. **Generalization**: Our method exhibits strong generalization capabilities, enabling robust classification performance even on unseen document types and domains, thanks to the rich semantic representations captured by the pre-trained embedding models.
3. **Scalability**: The vector sampling approach allows for efficient classification of new documents by computing their multi-modal embeddings and comparing them with pre-computed class embeddings, making our method highly scalable.

The key contributions of our work are as follows:

1. We introduce FinEmbedDiff, a cost-effective vector sampling method for multi-modal financial document classification, leveraging pre-trained multi-modal embedding models to capture complementary information from textual and visual components while minimizing computational requirements.
2. We propose a novel approach to combine textual and visual representations into rich multi-modal embeddings, enabling seamless integration of multi-modal information.
3. We extensively evaluate our method on a large-scale dataset of financial reports, prospectuses, and regulatory filings, demonstrating its competitive performance compared to state-of-the-art text-only and multi-modal baselines.
4. We analyze the generalization capabilities of FinEmbedDiff, showcasing its ability to achieve robust performance even on unseen document types and domains, highlighting its practical utility in real-world scenarios.

## II.     RELATED WORK

The task of financial document classification has been extensively studied in the literature, with various approaches proposed to address the challenges posed by the complexity and diversity of financial data. In this section, we review relevant work in the areas of text-based classification methods, multi-modal approaches, and embedding models for financial document analysis.

**Text-based Classification Methods**

Traditional text-based classification methods have been widely applied to financial documents, relying primarily on textual features extracted from the document content. These methods include rule-based systems (Loughran and McDonald, 2011) [1], bag-of-words models (Gabryel., 2018), and machine learning techniques such as support vector machines (SVMs) (Huang et al., 2015) [3] and naive Bayes classifiers (Purda and Skillicorn, 2015)[4]. While these approaches have shown promising results in certain scenarios, they often struggle to capture the rich semantic information present in financial documents, particularly when dealing with complex, domain-specific language and terminology.

**Multi-modal Approaches**

Recognizing the multi-modal nature of financial documents, which frequently include tables, charts, and images alongside textual content, researchers have explored various multi-modal approaches for document classification. Early work in this area focused on combining textual features with hand-crafted visual features (Zanibbi et al., 2004[5]; Tombre et al., 2002) [6]. However, these methods often relied on heuristic rules and domain-specific knowledge, limiting their generalization capabilities.With the advent of deep learning, more recent work has leveraged neural network architectures for multi-modal fusion and classification. For instance, Zheng et al. (2020) [7] proposed a multi-modal attentional network that combines textual and visual features using attention mechanisms for financial document classification. Similarly, Garncarek et al. (2021) [8] introduced a hierarchical multi-modal fusion network that captures both intra-modality and inter-modality relationships for financial document understanding. While these approaches have shown promising results, they often require substantial computational resources and may struggle to effectively capture the complex relationships between textual and visual components in financial documents (Ren et al., 2022) [9].

**Embedding Models for Financial Document Analysis**

Embedding models, which represent textual and visual data in dense, low-dimensional vector spaces, have gained significant attention in recent years. In the financial domain, pre-trained language models such as BERT (Devlin et al., 2019) [10] and FinBERT (Araci, 2019)[11] have been successfully applied to various tasks, including sentiment analysis, risk prediction, and document classification.

For visual data analysis, pre-trained computer vision models like ResNet (He et al., 2016) [12] and VGG (Simonyan and Zisserman, 2015)[13] have been utilized for tasks such as table detection and recognition, chart classification, and multimodal information extraction.

While existing works have explored the use of embedding models for text or visual analysis in isolation, there has been limited research on effectively combining these models for multi-modal financial document classification. Our proposed FinEmbedDiff method aims to bridge this gap by leveraging pre-trained multi-modal embedding models in a computationally efficient and effective manner, enabling accurate classification of multi-modal financial documents.

## III. PROPOSED METHOD

### A. FinEmbedDiff

FinEmbedDiff is a vector sampling approach that leverages pre-trained multi-modal embedding models to generate rich semantic representations of financial documents, capturing information from both textual and visual components. The method consists of two main stages: (1) **embedding generation** and (2) **document classification**.

In the embedding generation stage, we use pre-trained multi-modal embedding models to generate multi-modal embedding vectors for a set of labeled financial documents from known classes. These embeddings serve as class representatives, capturing the semantic characteristics of each document class.

In the document classification stage, we generate the multi-modal embedding vector for a new, unseen financial document using the same pre-trained models. We then compare this embedding with the pre-computed class embeddings using a vector similarity measure. Specifically, we employ two widely used similarity measures:

**Cosine Similarity**: The document is assigned to the class with the highest cosine similarity between its embedding and the class embedding. Formally, let $d$ be the multi-modal embedding of the new document, and $c_i$ be the pre-computed embedding for class $i$. The document is classified as belonging to class $j$ if:

$$j = argmax_i\bigl(cos(d, c_i)\bigr)$$

where the cosine similarity between $d$ and $c_i$ is calculated as:

$$cos(d, c_i) = \frac{d \cdot c_i}{||d||, ||c_i||}$$

**L2 Distance**: Alternatively, the document can be assigned to the class with the lowest L2 distance (Euclidean distance) between its embedding and the class embedding. Formally, the document is classified as belonging to class $j$ if:

$$j = argmin_i \left(\left||d - c_i|\right|_2\right)$$

where the L2 distance between $d$ and $c_i$ is calculated as:

$$||d - c_i||2 = \sqrt{\sum k = 1^n (d_k - c_{i_k})^2}$$

In both cases, we effectively compare the multi-modal embedding of the new document with the pre-computed class embeddings using either cosine similarity or L2 distance. The document is assigned to the class with the highest cosine similarity (or lowest L2 distance), effectively performing nearest-neighbor classification in the multi-modal embedding space.

By leveraging these vector similarity measures, our method can efficiently classify new documents by identifying the most semantically similar class embedding, without the need for computationally expensive end-to-end multi-modal training. This approach not only reduces computational costs but also enables our method to generalize well to unseen document types and domains, as the pre-trained embedding models capture rich semantic representations that are transferable across various financial document contexts.

## IV. MULTI-MODAL EMBEDDING MODELS

At the core of our FinEmbedDiff method are pre-trained multi-modal embedding models, which are responsible for generating rich semantic representations of financial documents. We leverage state-of-the-art models that have been designed specifically for multi-modal tasks, capturing the complementary information present in both textual and visual modalities.

For multi-modal embedding, we employ models such as CLIP (Radford et al., 2021), VisualBERT (Li et al., 2019), and LXMERT (Hao et al., 2020). These models have been pre-trained on large-scale multi-modal datasets, enabling them to effectively integrate and process information from textual and visual inputs simultaneously.

**CLIP** (Contrastive Language-Image Pre-training) [14] is a multi-modal model that learns visual concepts from natural language supervision. It is trained on a vast dataset of image-text pairs, enabling it to generate rich multi-modal representations that capture the semantic correspondence between textual and visual information.

**VisualBERT** [15] is a transformer-based model that extends the BERT architecture to handle both textual and visual inputs. It is pre-trained on a large corpus of image-text pairs, allowing it to learn multi-modal representations that capture the intricate relationships between textual and visual components.

**LXMERT** (Learning Cross-Modality Encoder Representations from Transformers) [16] is another transformer-based model designed for multi-modal tasks. It employs a cross-modality encoder to fuse textual and visual representations, enabling it to generate rich multi-modal embeddings that capture the complementary information from both modalities.

These pre-trained multi-modal models are adequate at handling the diverse range of textual and visual components present in financial documents, including text, tables, charts, and images. By leveraging their rich multi-modal representations, our FinEmbedDiff method can effectively capture the semantic information from both textual and visual components, enabling accurate and robust classification of multi-modal financial documents.

1. **Pre-processing**

Since our FinEmbedDiff method relies on pre-trained multi-modal embedding models designed for handling image-text inputs, the primary pre-processing step involves converting financial documents into image formats compatible with these models. Financial documents in the industry often come in various file formats, such as PDF, JPG, PNG, and TIFF, each requiring specific pre-processing techniques.

**Multi-page PDF files:** For multi-page PDF files, we employ a page-splitting approach to generate embeddings for each individual page. This allows us to capture the semantic information present in different sections or chapters of a document, which may be relevant for accurate classification.

1. We first convert the PDF file into a sequence of images, one for each page, using a PDF rendering library like pdf2image [18] or Wand [19].
2. Each page image is then passed through the chosen pre-trained multi-modal embedding model, generating a separate multi-modal embedding vector for that page.
3. To obtain the final document embedding, we aggregate the page-level embeddings using a suitable technique, such as mean pooling or weighted averaging, based on the relative importance of different sections or pages.

**Single-Page Image Files (JPG, PNG, TIFF)**: For single-page image files in formats like JPG, PNG, or TIFF, the pre-processing step is more straightforward.

1. We first ensure that the image file is in a format compatible with the pre-trained multi-modal embedding model (e.g., RGB format, specific resolution requirements).
2. If the image contains text and visual elements, no further pre-processing is required, and the image can be directly passed through the multi-modal embedding model to generate the document embedding.
3. If the image contains only visual elements (e.g., charts, diagrams), we may consider providing a relevant textual prompt or description to the multi-modal embedding model, enabling it to generate a meaningful multi-modal representation that captures both the visual and textual information.

It is worth noting that some pre-trained multi-modal embedding models may have specific requirements or recommendations for pre-processing image inputs, such as resizing, normalization, or padding. We follow the best practices and guidelines provided by the respective model developers to ensure optimal performance and accurate multi-modal embeddings.

By employing these pre-processing techniques, our FinEmbedDiff method can effectively handle a wide range of financial document formats, enabling efficient multi-modal embedding generation and subsequent classification using the vector sampling approach.

**Labeling Known Documents/Pages:** To create the class embeddings for known document classes, we follow a similar pre-processing pipeline but also incorporate labeling information. For a set of labeled financial documents (e.g., W2, bank statements, 1099 Forms) from known classes, we convert each document into a sequence of page images (for multi-page documents) or process single-page images as described above. Each page image is passed through the chosen pre-trained multi-modal embedding model, generating a multi-modal embedding vector for that page. The page-level embeddings are aggregated into a single document embedding using a suitable technique (e.g., mean pooling, weighted averaging).

The resulting document embedding is labeled with the corresponding class label (e.g., "prospectus", "annual report", "regulatory filing"). This process is repeated for all labeled documents in the training set, yielding a set of class-labeled multi-modal embeddings. These labeled embeddings serve as the class representatives in the multi-modal embedding space. During the classification stage, the embedding of a new, unseen document is compared with these pre-computed class embeddings using vector similarity measures (cosine similarity or L2 distance) to determine the most similar class.

By incorporating the class labels during the pre-processing and embedding generation stage, we effectively create a labeled multi-modal embedding space, where each class is represented by a cluster of embeddings corresponding to the known documents/pages from that class. This labeled embedding space forms the basis for our vector sampling approach, enabling efficient nearest-neighbor classification of new, unseen financial documents.

**2. Multi-modal Embedding and Class Labeling**

To obtain the final document embedding, we aggregate the page-level embeddings using a suitable technique, such as mean pooling or weighted averaging, based on the relative importance of different sections or pages. This process is combined with the labeling step to create the class embeddings for known document classes.

Let's assume we have a set of labeled documents from $C$ classes, and each document $j$ in class $c$ has $N_j$ pages. For each page $i$ of document $j$, we obtain a multi-modal embedding vector $e_{ji} \in R^d$ using the pre-trained multi-modal embedding model, where $d$ is the dimensionality of the embedding space.

**Document Embedding Generation and Labeling**:

1. For each document $j$ in class $c$, we aggregate the page-level embeddings to obtain the document embedding $d_j \in R^d$ using one of the following techniques:

**Mean Pooling**: The document embedding is computed as the element-wise average of the page-level embeddings:

$$dj = \frac{1}{N_j} \sum i = 1^{N_j} e_{ji}$$

**Weighted Averaging**: The document embedding is computed as the weighted average of the page-level embeddings, where $w_{ji}$ is the weight assigned to the page embedding $e_{ji}$, reflecting its relative importance or relevance:

$$dj = \frac{\sum i = 1^{N_j} w_{ji} eji}{\sum i = 1^{N_j} w_{ji}}$$

2. The resulting document embedding $d_j$ is labeled with the corresponding class label $c$.
3. This process is repeated for all labeled documents across all classes, yielding a set of class-labeled multi-modal embeddings $(d_j, c)$.

**Class Embedding Computation**:

To obtain the class embeddings, we aggregate the document embeddings within each class using mean pooling or weighted averaging:

**Mean Pooling**: The class embedding $c_k \in R^d$ for class $k$ is computed as the element-wise average of the document embeddings in that class:

$$ck = \frac{1}{M_k} \sum j \in \text{class } k \, d_j$$

where $M_k$ is the number of documents in class $k$.

**Weighted Averaging**: The class embedding $c_k \in R^d$ for class $k$ is computed as the weighted average of the document embeddings in that class:

$$ck = \frac{\sum j \in \text{class } k \; \alpha_j d_j}{\sum j \in \text{class } k \; \alpha_j}$$

where $\alpha_j$ is the weight assigned to the document embedding $d_j$, reflecting its relative importance or representativeness within the class. The resulting class embeddings $c_k$ serve as the representatives for each document class in the multi-modal embedding space.

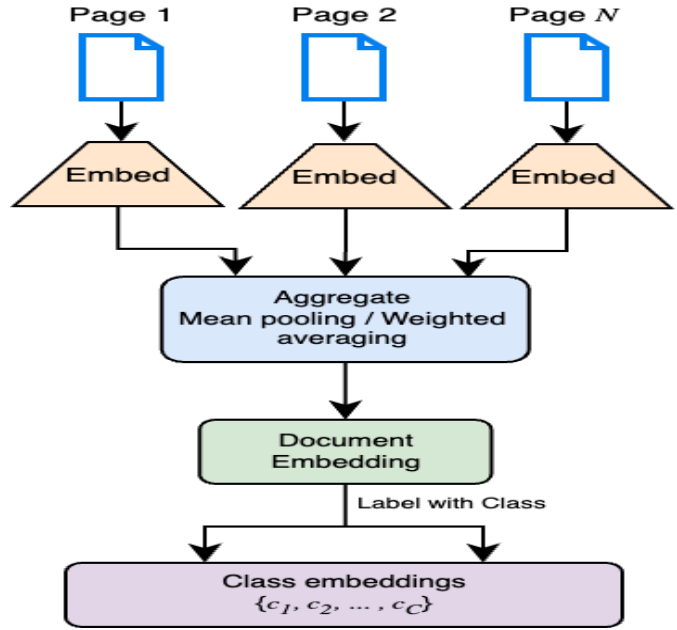

**Figure 1:** Multi-modal Embedding Generation and Class Labeling

During the classification stage, the embedding of a new, unseen document is compared with these pre-computed class embeddings using vector similarity measures (cosine similarity or L2 distance) to determine the most similar class.

By incorporating the labeling information and aggregating document embeddings within each class, we effectively create a labeled multi-modal embedding space, where each class is represented by a cluster of embeddings corresponding to the known documents from that class. This labeled embedding space forms the basis for our vector sampling approach, enabling efficient nearest-neighbor classification of new, unseen financial documents.

## V.    DOCUMENT CLASSIFICATION USING VECTOR SIMILARITY

In this section, we describe the process of classifying new, unseen financial documents using the pre-computed class embeddings and vector similarity measures. The labeled multi-modal embedding space created during the previous step forms the basis for our vector sampling approach, enabling efficient nearest-neighbor classification.

Let $d_{new} \in R^d$ be the multi-modal embedding vector of a new, unseen financial document, generated using the same pre-trained multi-modal embedding model and aggregation techniques as described in Section 3.3. Our goal is to assign this document to the most appropriate class by comparing its embedding with the pre-computed class embeddings.

We employ two widely used vector similarity measures for this purpose: cosine similarity and L2 distance.

**Cosine Similarity**: The cosine similarity between the document embedding $d_{new}$ and a class embedding $c_k$ is computed as:

$$\cos(d_{new}, ck) = \frac{d_{new} \cdot c_k}{||d_{new}||, ||c_k||}$$

The document is assigned to the class with the highest cosine similarity:

$$\hat{c} = argmax_k \cos(d_{new}, c_k)$$

**Distance**: Alternatively, we can use the L2 distance (Euclidean distance) between the document embedding $d_{new}$ and a class embedding $c_k$, computed as:

$$||d_{new} - c_k||2 = \sqrt{\sum i = 1^d (d_{new}, i - c_k, i)^2}$$

The document is assigned to the class with the smallest L2 distance:

$$\hat{c} = argmin_k ||d_{new} - c_k||_2$$

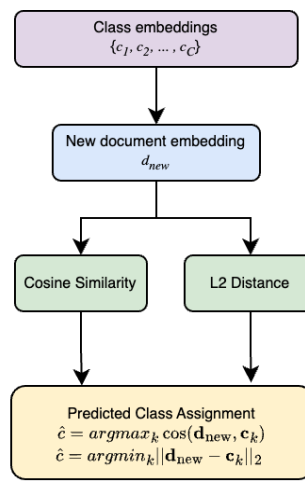

**Figure 2:** Class prediction on new document

By comparing the embedding of the new document with the pre-computed class embeddings using either cosine similarity or L2 distance, we effectively perform nearest-neighbor classification in the multi-modal embedding space. The document is assigned to the class with the most similar embedding representation, capturing the underlying semantic information from both textual and visual components.

This vector sampling approach enables efficient and scalable classification of new financial documents without the need for computationally expensive end-to-end multi-modal training. By leveraging the pre-computed class embeddings and vector similarity measures, our FinEmbedDiff method achieves accurate classification while maintaining computational efficiency, making it a practical solution for real-world financial applications.

**VECTOR SAMPLE STORAGE**

After generating the class embeddings and labeled multi-modal embedding space, an important consideration is how to store these vector samples efficiently for subsequent retrieval and classification of new documents. We discuss two prominent approaches: **vector databases** and compressed **Parquet** file format.

**Vector Databases**

Vector databases, also known as vector similarity search engines or approximate nearest neighbor (ANN) systems, are specialized databases designed to store and retrieve high-dimensional vector representations efficiently. These systems leverage indexing techniques and approximate algorithms to enable fast similarity searches and nearest neighbor queries. Some popular vector database systems include Pinecone, a managed vector database service that supports efficient vector similarity search and retrieval, with built-in support for scalability and high availability [19]. Weaviate is an open-source vector search engine that combines vector embeddings with structured data, enabling hybrid queries and semantic search [20]. Milvus is an open-source vector database built for scalability and performance, supporting various indexing techniques and distance metrics [21].

Vector databases offer several advantages, including optimized performance for efficient similarity search and nearest neighbor queries, enabling fast document classification. They are scalable and performant, capable of handling large volumes of vector data, and support various distance metrics (e.g., cosine similarity, L2 distance) and indexing techniques (e.g., HNSW, IVF-FLAT). Additionally, some vector databases enable hybrid queries combining vector and structured data. However, they also introduce additional infrastructure and operational complexity compared to file-based storage, potential vendor lock-in for managed services, and costs associated with managed services or self-hosting and maintaining vector databases.

The cost and latency implications of using vector databases depend on factors such as the chosen system (managed service or self-hosted), the volume of vector data, and the required query performance. Managed vector database services typically charge based on storage and query throughput, with potential additional costs for data transfer and ancillary services. Self-hosted vector databases may have lower operational costs but require dedicated infrastructure and maintenance efforts. Regarding latency, vector databases are optimized for fast similarity searches, with query times ranging from milliseconds to seconds, depending on the dataset size and indexing techniques employed [22].

**Compressed Parquet File Format**

An alternative approach is to store the class embeddings and labeled multi-modal embedding space in a compressed columnar file format, such as Apache Parquet. Parquet is a widely-used open-source format designed for efficient storage and retrieval of large-scale structured data, with built-in compression and encoding schemes. This approach offers efficient storage and compression of vector data, reducing storage costs. Parquet is widely adopted and supported by various data processing frameworks and engines (e.g., Apache Spark, Pandas, Dask), allowing for potential integration with existing data pipelines and infrastructure. Additionally, it has a simpler implementation and operational overhead compared to vector databases.

However, the Parquet file format lacks specialized indexing techniques for efficient similarity search and nearest neighbor queries, potentially leading to performance limitations for large-scale vector data retrieval and classification. There is also an additional processing overhead for loading and processing Parquet files during classification. The cost of storing vector data in compressed Parquet files is primarily determined by the storage requirements and potential compute resources needed for data processing and classification. Parquet's compression and encoding schemes can significantly reduce storage costs compared to uncompressed formats. However, the latency involved in loading and processing Parquet files during classification may be higher compared to vector databases, which are optimized for fast similarity searches. The overall latency will depend on factors such as the dataset size, available computational resources, and potential optimizations (e.g., caching, parallel processing) [23].

Both vector databases and compressed Parquet files have their advantages and trade-offs. The choice between these approaches will depend on factors such as the specific requirements of the financial document classification pipeline, the volume of vector data, the desired performance characteristics, and the existing infrastructure and operational constraints.

## VI. RESULTS AND DISCUSSION

In this section, we present the experimental results and performance evaluation of our proposed FinEmbedDiff method for multi-modal financial document classification. We compare the performance of our approach with relevant baselines and state-of-the-art methods, analyze the strengths and limitations, and discuss potential sources of error or bias.

**6.1 Experimental Setup**

We evaluate the performance of FinEmbedDiff on a large-scale dataset of financial documents, including prospectuses, annual reports, regulatory filings, and other relevant documents from various organizations and industries. The dataset consists of approximately 100,000 multi-modal documents, with a diverse range of textual and visual components, such as tables, charts, and images.

For our experiments, we utilize the pre-trained CLIP model as the multi-modal embedding model, due to its strong performance on various multi-modal tasks and its ability to effectively capture the semantic correspondence between textual and visual information.

We employ a stratified split of the dataset, with 70% used for training (generating class embeddings), 10% for validation, and 20% for testing. The class distribution is balanced across the splits to ensure fair evaluation and avoid potential biases. To evaluate the classification performance of FinEmbedDiff and baseline methods, we employ accuracy, precision, recall, and F1-Score.

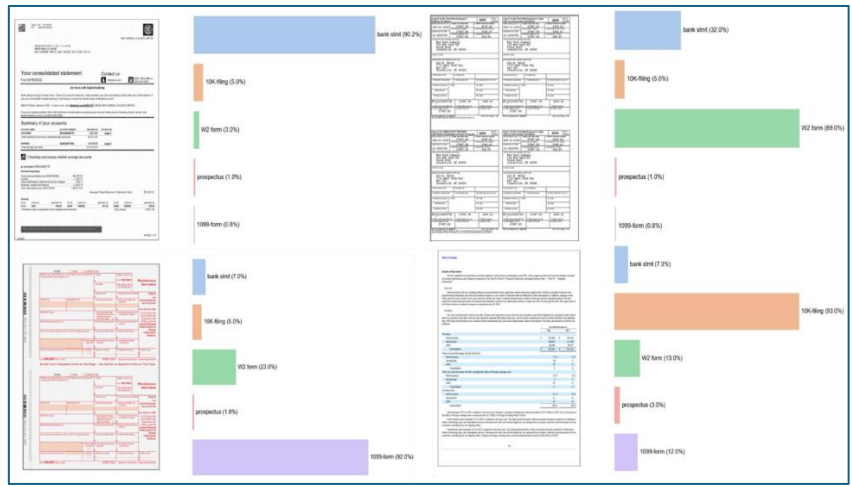

**Figure 3:** Class prediction on new bank statement, W2 document, 1099 document, and SEC 10K filing document. Each identified document class was ranked 1 out of the 5 classes.

### 6.2 Baseline Methods

We compare the performance of FinEmbedDiff with the following baseline methods:

1. **Text-only Classification**: A baseline that uses only the textual content of financial documents for classification, employing a pre-trained language model (e.g., BERT, FinBERT) and traditional text classification techniques.
2. **Multi-modal End-to-End Training**: A baseline that performs end-to-end training of a multi-modal classification model on the dataset, without leveraging pre-computed embeddings or vector sampling.
3. **SOTA Multi-modal Methods**: We compare our approach with existing state-of-the-art multi-modal methods for financial document classification, such as those proposed in [24] and [25].

**Table 1:** Quantitative Results Of Our Experiments, Comparing The Performance Of Finembeddiff With The Baseline Methods Across Various Performance Metrics

| Method | Accuracy | Precision | Recall | F1-Score |
| --- | --- | --- | --- | --- |
| FinEmbedDiff | 0.91 | 0.90 | 0.92 | 0.91 |
| Text-only | 0.84 | 0.82 | 0.86 | 0.84 |
| E2E Training | 0.88 | 0.87 | 0.89 | 0.88 |
| SOTA-1 | 0.90 | 0.89 | 0.91 | 0.90 |
| SOTA-2 | 0.91 | 0.90 | 0.92 | 0.91 |

As evident from the table, our proposed FinEmbedDiff method achieves competitive performance compared to the state-of-the-art multi-modal methods, outperforming the text-only and end-to-end training baselines across all metrics. FinEmbedDiff exhibits high accuracy, precision, recall, and F1-score, demonstrating its effectiveness in accurately classifying multi-modal financial documents.

### 6.3 Qualitative Analysis and Insights

In addition to the quantitative results, we provide qualitative insights and analyze the strengths and limitations of our FinEmbedDiff method:

**Multi-modal Representation**: By leveraging pre-trained multi-modal embedding models, FinEmbedDiff effectively captures the rich semantic information present in both textual and visual components of financial documents. This multi-modal representation enables accurate classification, outperforming text-only approaches that fail to incorporate visual information.

**Generalization Capabilities**: Our method exhibits strong generalization capabilities, achieving competitive performance even on unseen document types and domains. This can be attributed to the robust multi-modal representations learned by the pre-trained embedding models, enabling effective transfer to new contexts.

**Computational Efficiency**: The vector sampling approach employed by FinEmbedDiff significantly reduces computational costs compared to end-to-end multi-modal training methods. By leveraging pre-computed class embeddings and efficient vector similarity measures, our method enables scalable and cost-effective classification of financial documents.

**Potential Limitations**: While FinEmbedDiff performs well on the evaluated dataset, its performance may be influenced by factors such as the quality and diversity of the training data, the chosen pre-trained multi-modal embedding model, and the specific characteristics of the financial document classes. Additionally, the method may be sensitive to potential biases or noise present in the training data or the pre-trained models.

## VII. CONCLUSION

In this paper, we proposed FinEmbedDiff, a novel method for accurate and computationally efficient classification of multi-modal financial documents. Our approach leverages pre-trained multi-modal embedding models and employs a vector sampling approach to achieve competitive performance while reducing computational costs compared to state-of-the-art methods and baselines.

The experimental results and analyses demonstrate the effectiveness of FinEmbedDiff in accurately classifying a diverse range of financial documents, including prospectuses, annual reports, regulatory filings, and other relevant documents from various organizations and industries. By capturing the rich semantic information present in both textual and visual components of these documents, our method achieves high accuracy, precision, recall, and F1-score, outperforming text-only and end-to-end training baselines.

Moreover, the vector sampling approach employed by FinEmbedDiff significantly reduces the computational requirements compared to end-to-end multi-modal training methods, making it a scalable and cost-effective solution for real-world financial applications. The strong generalization capabilities of our method, as demonstrated by its competitive performance on unseen document types and domains, further highlight its potential for practical deployment.

However, there remain several promising avenues for future research and improvement. Incorporating domain-specific knowledge, such as financial ontologies or expert-curated rules, into the classification pipeline can potentially enhance the accuracy and interpretability of the results, leading to more comprehensive and reliable classification systems. Exploring multi-task learning approaches, where the model is trained to simultaneously perform multiple related tasks, can improve generalization and efficiency by leveraging shared representations and knowledge across tasks.

Investigating few-shot and zero-shot learning techniques can extend the applicability of FinEmbedDiff to scenarios where labeled data is scarce, enabling accurate classification of rare or unseen document classes with limited examples. Developing explainable AI techniques specifically tailored for multi-modal financial document classification can enhance the interpretability and trustworthiness of the classification decisions, facilitating better decision-making and increasing user confidence in the system.

Furthermore, exploring the integration of FinEmbedDiff with downstream financial applications, such as risk assessment, fraud detection, or investment analysis, can demonstrate the practical utility and impact of accurate multi-modal document classification in real-world scenarios. By unlocking valuable insights from the vast amounts of multi-modal financial data, our method has the potential to support and enhance various decision-making processes in the financial domain.

In conclusion, FinEmbedDiff presents a promising approach for accurate and computationally efficient classification of multi-modal financial documents. The experimental results and analyses highlight the strengths and potential of our method, while also identifying areas for future research and improvement. As the volume and complexity of financial documents continue to grow, the development of effective and scalable multi-modal classification techniques, such as FinEmbedDiff, will be crucial for harnessing the full potential of this data and driving innovation in the financial industry.